# Electrical manipulation of spin pumping signal through nonlocal thermal magnon transport

Yabin Fan[1*], Justin T. Hou[1], Joseph Finley[1], Se Kwon Kim[2], Yaroslav Tserkovnyak[3] and Luqiao Liu[1]

[1]*Microsystems Technology Laboratories, Massachusetts Institute of Technology, Cambridge, Massachusetts 02139, USA*

[2]*Department of Physics and Astronomy, University of Missouri, Columbia, Missouri 65211, USA*

[3]*Department of Physics and Astronomy, University of California, Los Angeles, California 90095, USA*

**Abstract**: We study the magnon transport in the nonlocal configuration composed of two Pt strips on top of yttrium iron garnet, with and without the presence of RF microwave generated by an on-chip antenna. We find that the spin-Hall induced thermal magnon heating/cooling, the Oersted field as well as the Joule heating generated by the a.c. current in the Pt injector can significantly influence the spin-pumping signal measured by the Pt detector in the presence of RF microwave, forcing the spin-pumping voltage to show up in the first and second harmonic signals in the nonlocal magnon transport measurement. These results indicate that nonlocal magnon transport configuration can serve as a structure to electrically detect and manipulate the spin-pumping signal. Furthermore, certain caution is needed when studying the interplay between incoherent magnon and coherent magnon spin transport in the nonlocal transport configuration, since the change in microwave-induced spin-pumping voltage can overwhelm the incoherent magnon transport signals.

*Electronic mail: yabinfan@mit.edu

Recently, the incoherent magnon spin current generated electrically or thermally in magnetic insulators [1-3,4,5], compensated ferrimagnets[6,7] and antiferromagnetic insulators [8] have drawn great attention, due to their potential applications in miniaturized magnonic devices. The incoherent magnons can be generated through spin-Hall effect (SHE) in heavy metals [1,2,5], *e.g.*, in the Pt/yttrium iron garnet (YIG) structure by passing a charge current in the Pt layer, which creates a gradient in the magnon chemical potential in YIG [9]. They can also be generated through the process called spin-Seebeck effect (SSE) in the magnetic insulator [3,7] when a temperature gradient is present. These incoherent magnons have short wavelength at room temperature [10] since they have energy up to $k_\text{B}T/h \sim$ 6THz, where the exchange energy dominates. These short-wavelength, high-energy incoherent magnons could offer the avenue to miniaturize magnonic devices, as they have a diffusion length up to 10μm [1] and they are electrically controllable [11,12].

Besides incoherent magnons, the coherent magnons have been well-studied in the past. The coherent magnons can be generated by RF microwave at the ferromagnetic resonance (FMR) state [13,14,15], and they have well-defined frequency and long wavelength, as they can propagate on the centimeter scale. Therefore, they have been anticipated to have rich applications in magnonic logic devices [14,15,16]. The microwave generated coherent magnons usually fall into the GHz frequency region, where the magnetic dipole interaction dominates. Moreover, spin-orbit torque can also be employed to generate coherent magnons [17]. Naturally, it becomes important to investigate the interplay between coherent and incoherent magnons for their concurrent applications in magnonic devices, since they have different properties and they are generated and controlled by different means. However, despite large efforts, their interplay has still been under debate [15,18,19]. One natural way to investigate the interplay between them would be to study the electrically/thermally induced incoherent magnon transport in the nonlocal transport configuration in the presence of RF microwave [20]. We will show in the following that in the presence of RF microwave, the nonlocal magnon transport configuration composed of two Pt strips on top of magnetic insulator can pick up the spin-pumping signals caused by the microwave, which could overwhelm the incoherent magnon transport signals in this configuration. Moreover, our work also indicates that the spin-pumping signal can be manipulated electrically by nonlocal thermal magnon transport, offering a remote control of coherent magnetic dynamics.

To begin with, we have grown 60nm YIG films on top of the gadolinium gallium garnet (GGG) (110) substrate by magnetron sputtering [21], followed by rapid thermal annealing in oxygen atmosphere. To characterize the film quality, we have performed FMR measurement by placing the sample on top of a coplanar waveguide (CPW) to measure the RF microwave absorption while sweeping the in-plane magnetic field, as schematically shown in Fig. 1(a) inset. A typical FMR data at 6GHz microwave is plotted in Fig. 1(a). By fitting the FMR absorption spectrum with Lorentzian formula [22], we can obtain the resonance field $H_{\text{res}}$ and linewidth $\Delta H$ of the YIG film at the specific frequency. In Fig. 1(b), we show the obtained $H_{\text{res}}$ and $\Delta H$ under different frequencies. By fitting $\Delta H$ versus frequency with a linear curve [23], we get the damping factor $\alpha = 1.8 \times 10^{-4}$.

To further test the YIG film quality for incoherent magnon transport, we have fabricated Pt patterns on top by photolithography, as shown in Fig. 1(c) inset, to probe the nonlocal magnon transport. The pattern consists of two Pt bars (width:6μm, length:100μm, thickness:10nm) with square contact pads for wire bonding. The two Pt bars have a separation of 12μm (center to center). The idea is to use one of the two Pt bars as an injector and the other as a detector, to generate incoherent magnons by the SHE and the Joule heating in the Pt injector, and to measure the incoherent magnon signal by 1st and 2nd harmonic signals in the Pt detector, as has been reported previously [1,2]. We carried out the nonlocal magnon transport experiment by sending an a.c. current $I_{ac} = 16\text{mA}$ (r.m.s.) at 15.5Hz in the injector, and meanwhile measuring the 1st and 2nd harmonic voltages in the Pt detector bar when rotating the external field $H = 300\text{Oe}$ in-plane. As shown in Fig. 1(c), while the 2nd harmonic voltage shows $\cos(\theta)$ behavior, resulted from the inverse spin-Hall effect (ISHE) converting the thermally generated magnons to voltage signal, the 1st harmonic signal does not show explicit angle dependence, indicating that the SHE-induced magnons arriving at the detector is too weak to be detected, presumably due to the relatively high damping factor and large bar-to-bar separation on our film compared with previous reports [1,2].

To probe the RF microwave influence on the non-local measurement, we fabricated an on-chip antenna beside the two Pt bars in order to apply RF microwave, as schematically shown in Fig. 2(a). In this setup, a constant 4GHz 25dBm microwave generated by a microwave synthesizer is sent into the antenna, and the nonlocal magnon transport measurement is simultaneously carried out using a lock-in amplifier. As shown in Fig. 2(b), when we apply a 5mA (r.m.s.) 15.5Hz a.c. current through the Pt injector, we detected some dips in the 1st harmonic nonlocal voltage $V^{1w}$ on the Pt detector at the YIG resonance field positions during sweeping magnetic field along *x*-direction. These dips scale linearly with the a.c. current applied through the injector in the low current regime ($I_{ac} \leq 5\text{mA}$). To understand the mechanism of these dips, we carried out another experiment with a different measurement configuration.

In this second experiment, we apply amplitude modulated (50% modulation) microwave onto the antenna, and at the same time measure the voltage at the Pt detector bar with a lock-in amplifier at the same modulation frequency [24] when sweeping the magnetic field. This is in nature a spin pumping experiment configuration [25], where the RF spin pumping is induced by the on-chip microwave antenna. In the meanwhile, we can still apply d.c. current through the injector Pt bar of Fig. 2(a), which can provide additional influence onto the detected signal. As the current in the injector branch is at a different frequency from the detected signal (d.c. vs amplitude-modulation frequency), it does not directly contribute to the detected spin pumping voltage. As shown in Fig. 2(c), we can indeed measure the spin-pumping voltages $V_{sp}$ at the resonance field when the d.c. current is 0mA and ±5mA in the injector bar. Compared with the case $I_{dc}$=0mA, the spin-pumping voltage measured when $I_{dc}$=±5mA shows a smaller value and the resonance field moves towards a higher field, presumably due to the Joule heating effect from the injector which alters the material or interface property at the Pt detector area. In order to compare the 1$^{st}$ harmonic nonlocal signal measured in Fig. 2(b) with these data, we did subtraction between the $V_{sp}(I_{dc}$=+5mA) and the $V_{sp}(I_{dc}$=-5mA) data and present the result in Fig. 2(d). Since the 1$^{st}$ harmonic nonlocal voltage $V^{1w}$ in Fig. 2(b) is proportional to the current (rather than current's absolute value), it should correlate with the voltage difference between $V_{sp}(I_{dc}$=+5mA) and $V_{sp}(I_{dc}$=-5mA). As shown in Fig. 2(d), the differential $\Delta V_{sp}$ indeed shows some dip features around the resonance field, and in order to understand these features, we take a closer look at the spin-pumping signals in the following.

In Fig. 3(a), we plot the zoom-in image of the spin-pumping voltage for $I_{dc}$=±5mA of Fig. 2(c) in the positive resonance field regime. We find for both $I_{dc}$=+5mA and $I_{dc}$=-5mA, the spin-pumping peak splits into several peaks and the linewidth is greatly broadened, which could be attributed to the Joule heating effect that can reduce the YIG magnetization and introduce temperature inhomogeneity in the Pt detector area. More importantly, we notice that the $I_{dc}$=-5mA spin-pumping curve shows a larger linewidth broadening in the resonance high-field edge than the $I_{dc}$=+5mA curve. This difference is most possibly due to the SHE in the Pt injector: when $I_{dc}$=-5mA, the SHE in the Pt injector can result in spin accumulation pointing along *x*-direction at the Pt/YIG interface, as shown in Fig. 3(b) lower panel, and these spins can generate incoherent magnons in YIG by interfacial spin scattering [1], which can be regarded as a magnon "heating" process in addition to the Joule heating effect [11,19]; on the other hand, when $I_{dc}$=+5mA, the SHE in the Pt injector can result in spin accumulation pointing along -*x* -direction at the Pt/YIG interface, as shown in Fig. 3(b) upper panel, and these spins will absorb incoherent magnons in YIG by interfacial spin scattering [1], which can be regarded as a magnon "cooling" process [11,19]. The SHE induced heating/cooling leads to the difference in linewidth broadening in the $I_{dc}$=±5mA spin-pumping curves. Interestingly, the linewidth broadening difference is more obvious at the resonance high-field edge than the resonance low-field edge in Fig. 3(a), which could arise partially due to the different Oersted field generated by $I_{dc}$=±5mA in the injector [23]. Significantly, we observe reversed linewidth broadening difference for the $I_{dc}$=±5mA spin-pumping curves in the negative resonance field regime, as plotted in Fig. 3(c), which is expected since the SHE induced heating/cooling effect switches sign when the YIG magnetization is reversed (Fig. 3(d)). The difference in the $I_{dc}$=±5mA spin-pumping curves, after subtraction, gives rise to the dip features in Fig. 2(d). By comparing results in Fig. 2(b) and Fig. 2(d), it seems the dips appearing at the resonance field in the 1$^{st}$ harmonic nonlocal voltage $V^{1w}$ could be due to the electrical manipulation of the spin-pumping voltage through SHE induced heating/cooling and the Oersted field effect. More quantitatively, the dip in Fig. 2(b) versus the a.c. current/RF power ratio is, $(V_{dip}^{1w}/I_{ac})/P_{RF} = 25.3\ \mu V \cdot A^{-1} \cdot W^{-1}$, while the dip in $\Delta V_{sp}$ (Fig. 2(d)) versus the d.c. current/modulated RF power ratio reads, $(\Delta V_{sp}^{dip}/\Delta I_{dc})/\Delta P_{RF} = 29.5\ \mu V \cdot$ $A^{-1} \cdot W^{-1}$, suggesting their consistency. Here, $\Delta P_{RF}$ is the r.m.s value of the modulated RF power during the spin-pumping experiment shown in Fig. 2(c).

Now, we will analyze the 2nd harmonic nonlocal voltage $V^{2w}$. To check the universality of the microwave influence, this time we apply a 3GHz 20dBm microwave through the antenna. As shown in Fig. 4(a), in order to clearly measure the heating-induced magnon current flow from the injector to the detector, we apply a relatively large current, $I_{ac}$=16mA (r.m.s.), through the injector. Both 2nd harmonic signals, measured with the constant RF microwave and with no RF microwave applied in the antenna, are presented in Fig. 4(a). It can be clearly seen that, in addition to the SSE-induced 2nd harmonic voltage step at around $H_x$=0Oe [1], when the constant microwave is applied, a dip (peak) is developed at the positive (negative) FMR resonance field position. The dip (peak) value increases with the $I_{ac}$ current and gets saturated when $I_{ac} \geq$12mA. Again, we carry out the amplitude modulation spin-pumping experiment in the presence of different d.c. current applied through the Pt injector, as shown in Fig. 4(b). We can see that when the d.c. current is large, $I_{dc}$=±14mA, the spin-pumping voltage signal $V_{sp}$ is greatly suppressed and the linewidth is significantly broadened. As we have discussed, this could be due to the heating-induced modification of the YIG property and/or the Pt/YIG interfacial property (such as the spin-mixing conductance) in the Pt detector area. Further measurement of the Pt detector resistance change indicates the local temperature can be as high as 97°C when $I_{dc}$=±14mA in the injector. Since the 2nd harmonic nonlocal voltage detected in Fig. 4(a) is proportional to $I^2$, it is natural to correlate it with the $V_{sp}(I_{max})$-$V_{sp}(I$=0mA) data, as this difference in the spin-pumping voltage could in principle show up in the 2nd harmonic nonlocal voltage $V^{2w}$. So, in Fig. 4(c) we plot the subtraction between the $V_{sp}(I_{dc}$=+14mA) and the $V_{sp}(I_{dc}$=0mA) data obtained from Fig. 4(b). Clearly, the $\Delta V_{sp}$ data in Fig. 4(c) shows a close correlation with the data in Fig. 4(a), suggesting that the heating-induced modulation of the spin-pumping voltage signal could show up in the 2nd harmonic nonlocal voltage. More quantitatively, the $\Delta V_{sp}$ peak versus the d.c. current square/modulated RF power ratio in Fig. 4(c) reads, $(\Delta V_{sp}^{peak}/\Delta I_{dc}^2)/\Delta P_{RF} = 9.2$ mV · A$^{-2}$ · W$^{-1}$, while the $V^{2w}$ peak in Fig. 4(a) versus the a.c. current square/RF power ratio is, $V_{peak}^{2w}/(I_{ac}^2/\sqrt{2})/P_{RF} = 7.5$ mV · A$^{-2}$ · W$^{-1}$. The consistency in numbers obtained from the two experiments again confirms the proposed scenario, while the small discrepancy could be due to the difference in d.c. and a.c. heating efficiency.

In summary, in the nonlocal magnon transport configuration, with the application of RF microwave, the spin-pumping signal generated on the Pt detector can be modulated by the spin-Hall induced magnon heating/cooling, the Oersted field as well as the Joule heating generated by the a.c. current in the Pt injector, forcing the spin-pumping voltage to show up in the 1st and 2nd harmonic nonlocal voltage signals. It will be valuable to carefully calibrate such effect in the experiment during studying the interplay between the coherent and incoherent magnon transport, since this electrical- or heating- modulated spin pumping voltage may overwhelm the nonlocal incoherent magnon transport signals detected by the Pt detector. This problem originates in the fact that the ISHE in Pt detector cannot differentiate the spin-pumping voltage and the incoherent magnon spin signal, when they are both being modulated by the a.c. current in the injector. Optical method, such as the Brillouin light scattering (BLS) [26], might be a powerful tool to study the interplay between coherent and incoherent magnon transport, since BLS can probe magnons operating at different frequencies individually, and thus could potentially differentiate coherent and incoherent magnon signals in experiment.

See the supplementary material for several supplemental experiments that were conducted. Supplementary material section 1 presents estimation of the heating effect in the system and its influence on the spin-pumping. Section 2 shows the subtraction between two spin-pumping experiments measured under the same condition to rule out possible artifacts during data subtraction. Section 3 illustrates the current dependence of the 1$^{st}$ and 2$^{nd}$ harmonic nonlocal voltage dips under constant RF microwave.

This research was partially supported by National Science Foundation through the Massachusetts Institute of Technology Materials Research Science and Engineering Center DMR–1419807 and AFOSR under award FA9550-19-1-0048.

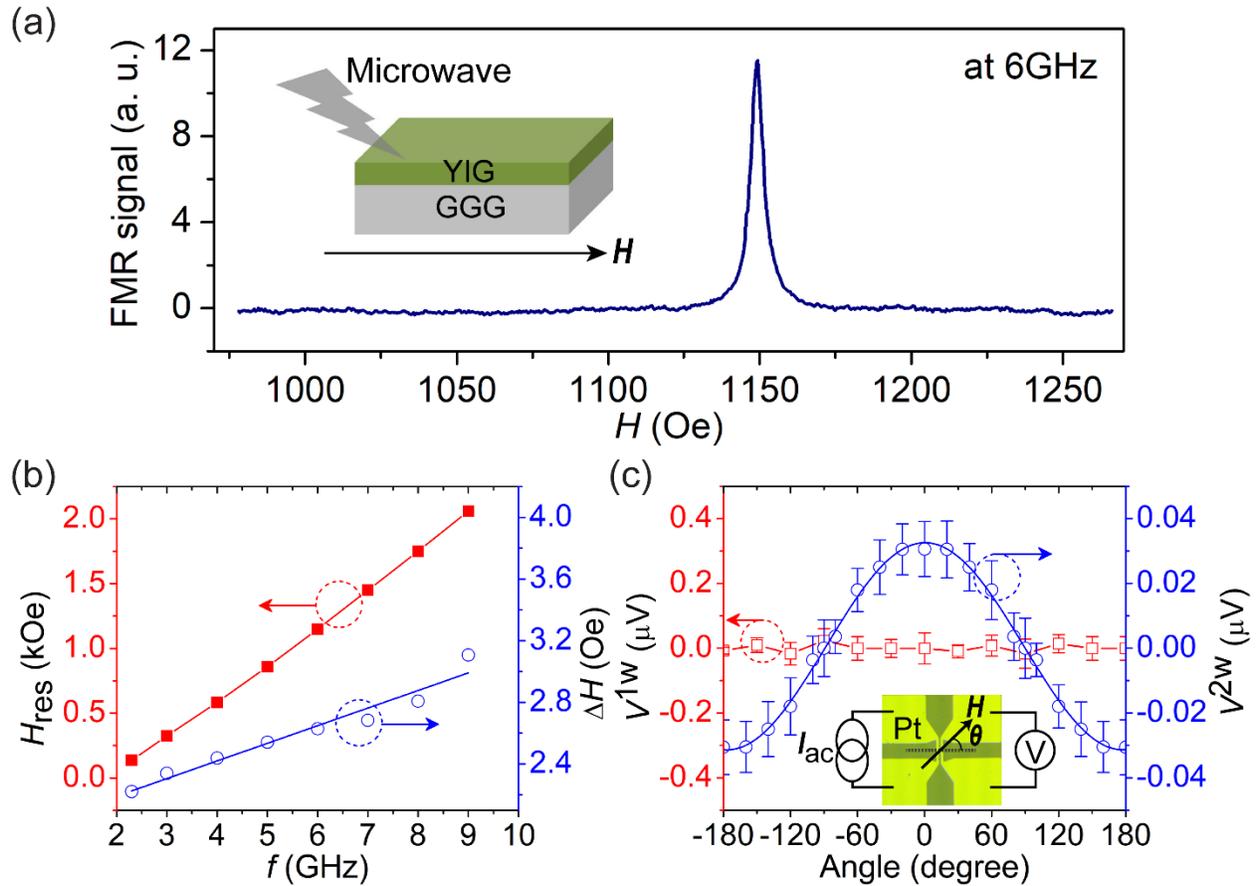

**Figure 1.** Material property measured by ferromagnetic resonance and non-local transport measurements. (a) Ferromagnetic resonance measurement on the GGG(substrate)/YIG(60nm) sample at 6GHz when sweeping the in-plane magnetic field. (b) The resonance field $H_{res}$ and the linewidth $\Delta H$ of the GGG/YIG(60nm) film at different microwave frequencies. Blue line is linear fit of $\Delta H$ versus $f$. (c) The nonlocal first and second harmonic transport measurement in the two Pt bars on top of the GGG/YIG(60nm) film, when the in-plane field ($H$=300Oe) rotates in the film plane. The a.c. current applied in the Pt injector is 16mA (r.m.s.) at 15.5Hz. The two Pt bars are 6μm wide, 100μm long and 10nm thick. They have a separation of 12μm (center to center). Measurement is carried out by a lock-in technique.

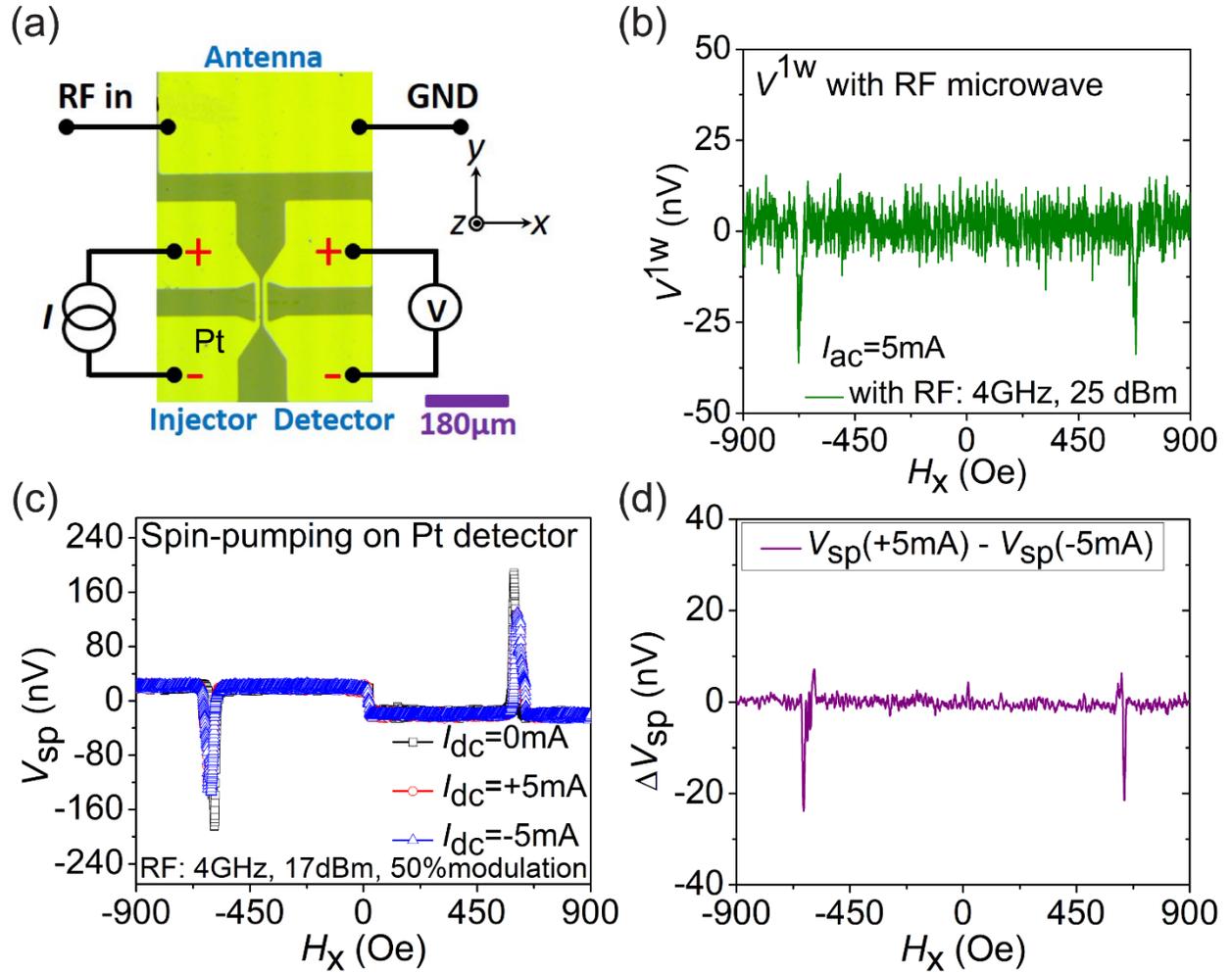

**Figure 2.** (a) Schematic of the experimental setup and the measurement scheme. The antenna, injector and detector are made of 10nm thick Pt films on top of GGG(substrate)/YIG(60nm) film. The injector and detector are both 6µm wide and 100µm long. The separation between them is 12µm (center to center). (b) The first harmonic nonlocal voltage measured by the Pt detector in the presence of 5mA (r.m.s.) 15.5Hz a.c. current injected in the Pt injector bar, when a constant 4GHz 25dBm microwave signal is sent into the antenna. (c) The spin pumping voltages measured by the detector Pt bar when different d.c. currents flow in the injector bar. The spin-pumping is measured by a lock-in technique with the microwave output power (17dBm at 4GHz) modulated by the lock-in via amplitude modulation (50% modulation). (d) The change in the $V_{sp}$ signal by subtraction between the $V_{sp}(I_{dc}=+5mA)$ and the $V_{sp}(I_{dc}=-5mA)$ data measured in (c).

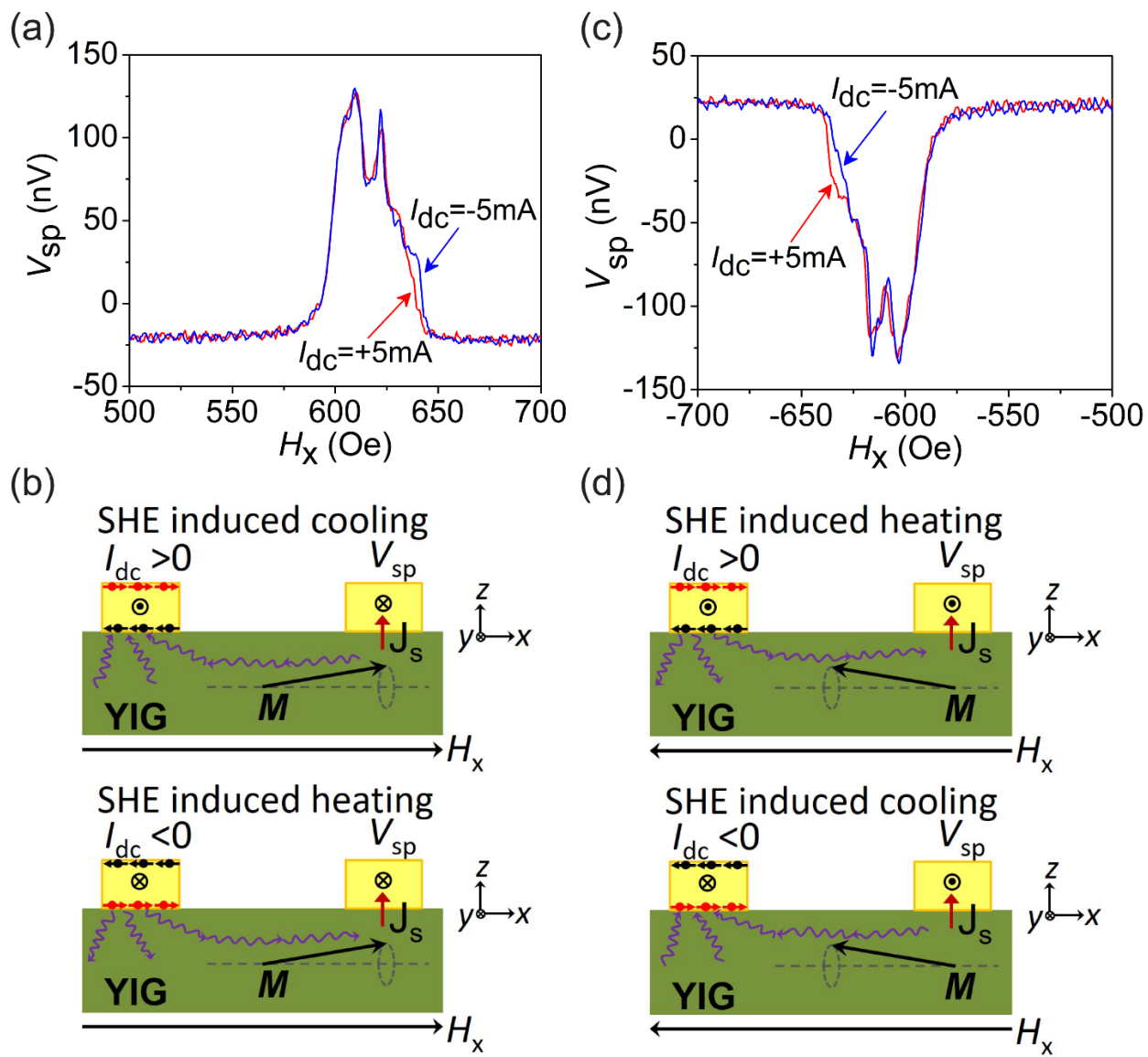

**Figure 3.** Zoom-in image of the spin pumping voltages measured by the detector Pt bar when different d.c. currents (±5mA) flow in the injector bar for the positive (a) and negative (c) external field resonance regime. The spin-pumping is measured by a lock-in technique with the microwave output power (17dBm at 4GHz) modulated by the lock-in via amplitude modulation (50% modulation). (b) and (d) show the SHE induced magnon heating and cooling effect when different d.c. currents flow in the injector bar under positive (b) and negative (d) external magnetic field, respectively.

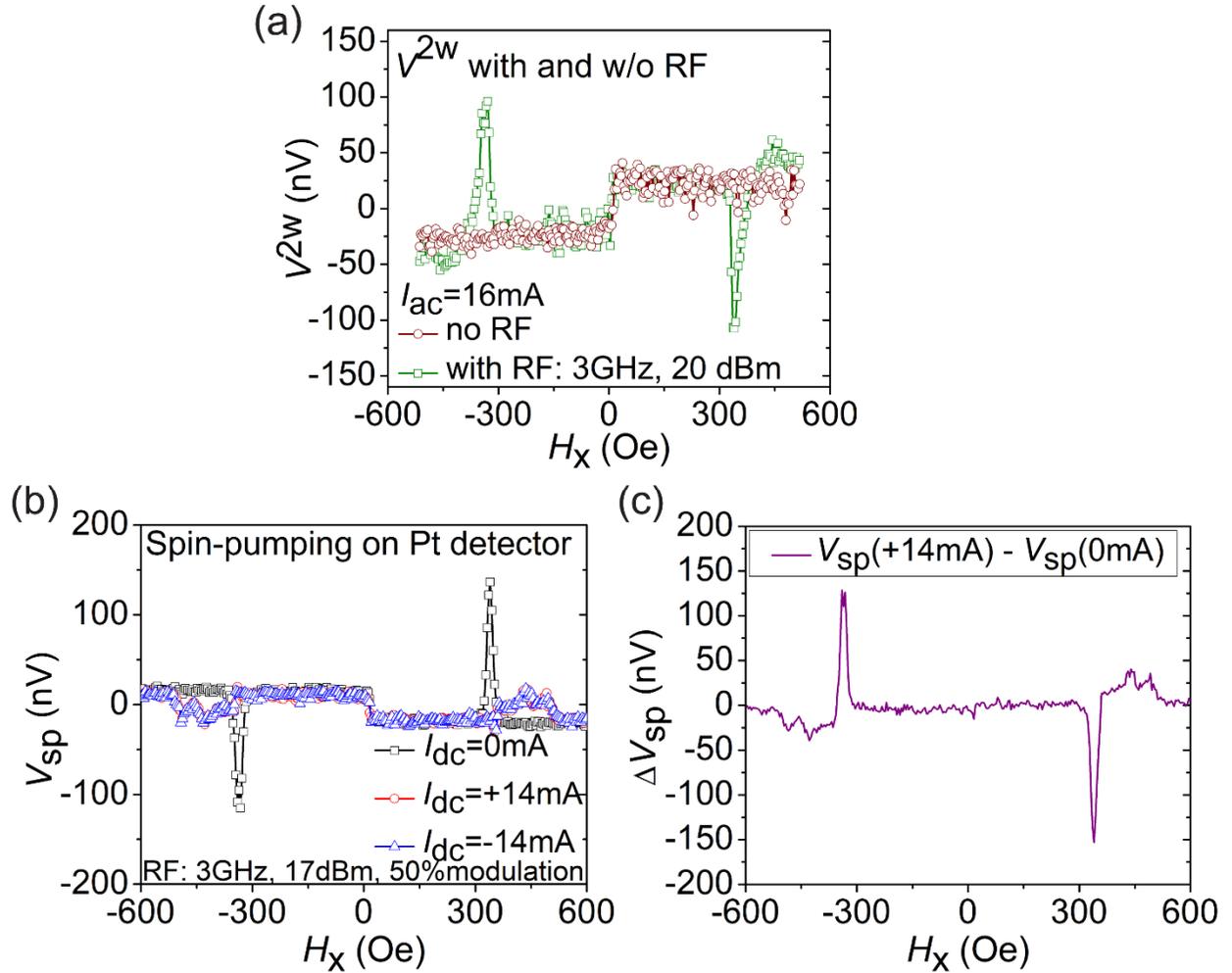

**Figure 4.** (a) The second harmonic nonlocal voltage measured by the Pt detector in the presence of 16mA (r.m.s.) 15.5Hz a.c. current injected in the Pt injector bar, when a constant 3GHz 20dBm microwave signal is sent into the antenna. The second harmonic voltage measured with no RF microwave applied is also presented as a reference. (b) The spin pumping voltages measured by the detector Pt bar when different d.c. currents flow in the injector bar. The spin-pumping is measured by a lock-in technique with the microwave output power (17dBm at 3GHz) modulated by the lock-in via amplitude modulation (50% modulation). (c) The change in the $V_{sp}$ signal by subtraction between the $V_{sp}(I_{dc}=+14mA)$ and the $V_{sp}(I_{dc}=0mA)$ data measured in (b).

# Supplementary Material for

# Electrical manipulation of spin pumping signal through nonlocal thermal magnon transport

Yabin Fan[1], Justin T. Hou[1], Joseph Finley[1], Se Kwon Kim[2], Yaroslav Tserkovnyak[3] and Luqiao Liu[1]

[1]*Microsystems Technology Laboratories, Massachusetts Institute of Technology, Cambridge, Massachusetts 02139, USA*

[2]*Department of Physics and Astronomy, University of Missouri, Columbia, Missouri 65211, USA*

[3]*Department of Physics and Astronomy, University of California, Los Angeles, California 90095, USA*

## Estimation of the heating effect in the system and its influence on the spin-pumping

In order to check the heating effect, we have measured the heating in the system using the following method. First, we measure the resistance of the Pt detector bar, $R_{\text{Detector}}$, as a function of the d.c. current applied in the Pt injector, $I_{\text{Injector}}$. As shown in Fig.S1(a), the $R_{\text{Detector}}$ increases dramatically with $I_{\text{Injector}}$, indicating a significant Joule heating effect by the current in the Pt injector bar. Next, we measure the $R_{\text{Detector}}$ as a function of the substrate temperature by heating the sample with an external heater. Here, no electrical current is applied through the Pt injector. As shown in Fig.S1(b), we get almost a linear dependence of $R_{\text{Detector}}$ versus temperature $T$. By comparing (a) and (b), we can estimate the temperature of the Pt detector bar area as a function of the d.c. current in the Pt injector (without external heater), as summarized in Fig.S1(c). In Fig.S1(d), we present the spin-pumping experiment detected by the Pt detector bar when different d.c. currents (0mA, +10mA) are applied in the Pt injector. From Fig.S1(c), we could get that the Pt detector area temperature is around 58 °C when $I_{\text{Injector}}$=+10mA. To crosscheck the heating effect on the spin-pumping spectrum, we also performed spin-pumping experiment at the ($I_{\text{Injector}}$=0mA, $T$=58 °C) condition by heating up the substrate via external heater, and the result is also plotted in Fig.S1(d). Compared with the $V_{\text{sp}}(I_{\text{dc}}$=0mA) spectrum (with no external heater), the $V_{\text{sp}}(I_{\text{dc}}$=0mA, $T$=58 °C) spin-pumping peaks shift towards higher field values, comparable with the $V_{\text{sp}}(I_{\text{dc}}$=+10mA) spectrum. We also observed that the $V_{\text{sp}}(I_{\text{dc}}$=0mA, $T$=58 °C) spin-pumping peak value is smaller than the $V_{\text{sp}}(I_{\text{dc}}$=0mA) spin-pumping peak, and the linewidth of $V_{\text{sp}}(I_{\text{dc}}$=0mA, $T$=58 °C) is a bit larger than that of the $V_{\text{sp}}(I_{\text{dc}}$=0mA) spectrum, presumably because the substrate heating increases the damping factor of the YIG film. Interestingly, the linewidth broadening for the $V_{\text{sp}}(I_{\text{dc}}$=+10mA) spin-pumping peak seems to be larger than the $V_{\text{sp}}(I_{\text{dc}}$=0mA, $T$=58 °C) peak, which is most likely due to the inhomogeneous Joule heating of the Pt detector area by the d.c. current in the Pt injector for the $V_{\text{sp}}(I_{\text{dc}}$=+10mA) case. In the $V_{\text{sp}}(I_{\text{dc}}$=0mA, $T$=58 °C) experiment, the substrate has been uniformly heated up.

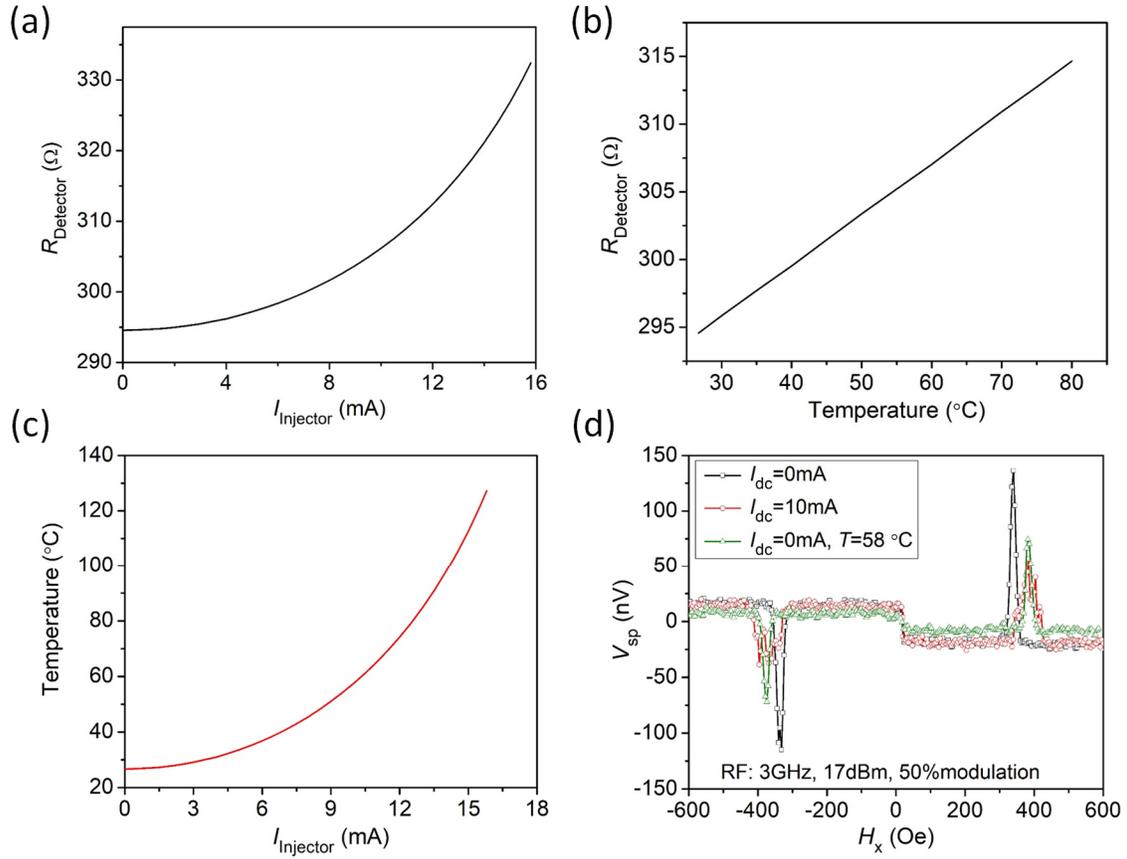

Figure S1. (a) The resistance of the Pt detector bar as a function of the current in the Pt injector bar. (b) The resistance of the Pt detector bar as a function of substrate temperature. (c) The extracted Pt detector bar area temperature versus the current in the Pt injector bar. (d) Spin-pumping voltage detected by the Pt detector bar when different d.c. currents (0mA, +10mA) are applied in the Pt injector. The spin-pumping signal detected with heated substrate (58 °C) by external heater is also shown as a comparison.

## Double-check of the spin-pumping experiment on the Pt detector in the presence of d.c. current in the Pt injector to rule out possible artifacts

We have carried out the spin-pumping experiment twice on the Pt detector when $I_{dc} = +5$mA is applied on the Pt injector, and the data subtraction is presented in the following figure, Fig.S2, to rule out possible artifacts that can be caused during data subtraction. In Fig.S2(a), we can see that the $V_{sp}$(+5mA, Test1) and $V_{sp}$(+5mA, Test2) data overlap with each other quite well. After subtraction, as shown in Fig.S2(b), we do not observe any peak or dip signals. So, the peaks and dips we have observed in Fig.2(d) and Fig.4(c) in the main text are real signals, and they are due to the modification of the spin-pumping spectrum by different currents in the Pt injector.

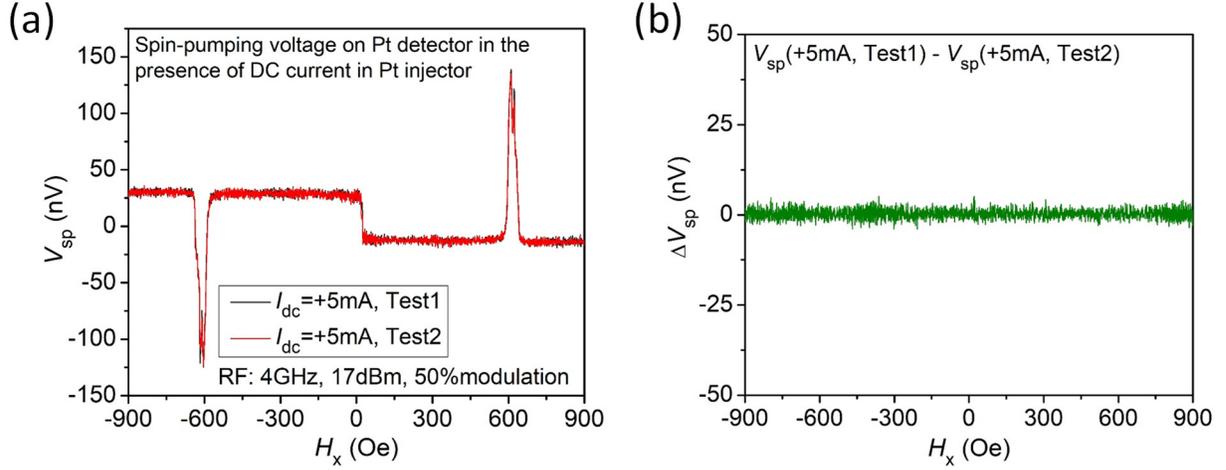

Figure S2. (a) Spin-pumping voltage measured on the Pt detector in the presence of d.c. current (+5mA) in the Pt injector at two different times, Test1 and Test2. (b) The subtraction between $V_{sp}$ (+5mA, Test1) and $V_{sp}$ (+5mA, Test2).

## Current dependence of the first and second harmonic nonlocal voltage dips

We use different d.c. current in Fig.4 and Fig.2 in the main text is because the 1$^{st}$ harmonic nonlocal voltage dip $V^{1w}$ measured in Fig. 2(b) has different current dependence compared with the 2$^{nd}$ harmonic nonlocal voltage dip $V^{2w}$ measured in Fig 4(a), at the resonance magnetic field position. In the following, we summarize the $V_{dip}^{1w}$ and $V_{dip}^{2w}$ dependence on the current in Fig. S3(a-b). The 1$^{st}$ harmonic nonlocal voltage dip $V_{dip}^{1w}$ shows a linear dependence on the current value in the small current regime, but it reaches the maximum value when $I_{ac}$=5mA in the injector. Further increase in the current would lead to decrease of the $V_{dip}^{1w}$ value. The reason is because $V_{dip}^{1w}$ is due to the spin-Hall effect (SHE)-induced shifting of the spin-pumping spectrum, and larger current will result in larger SHE. But the current can also cause broadening of the spin-pumping peak through Joule heating effect. When the current is larger than 5mA in the Pt injector, the significant linewidth broadening of the $V_{sp}$ peak will gradually smear out the shifting effect caused by the SHE. This is the reason why at $I_{ac}$=5mA, the 1$^{st}$ harmonic nonlocal voltage dip $V_{dip}^{1w}$ has reached the maximum value. Since the SHE-induced shifting of the spin-pumping spectrum is quite small, we have used higher RF power to optimize the signal-to-noise ratio in Fig.2(b) in the main text.

For the 2$^{nd}$ harmonic nonlocal voltage dip $V_{dip}^{2w}$ at the positive resonance magnetic field position, as shown in Fig.4(a) in the main text, it is primarily due to the suppression of the spin-pumping peak by the Joule heating effect. In other words, larger current will lead to more obvious effect. Indeed, as plotted in Fig.S3(b), the $V_{dip}^{2w}$ value increases with the $I_{ac}$ current applied through the Pt injector, and it reaches saturation value when $I_{ac}$ is above 12mA. So, we used quite large current value in Fig.4 in the main text in order to optimize the $V_{dip}^{2w}$ signal.

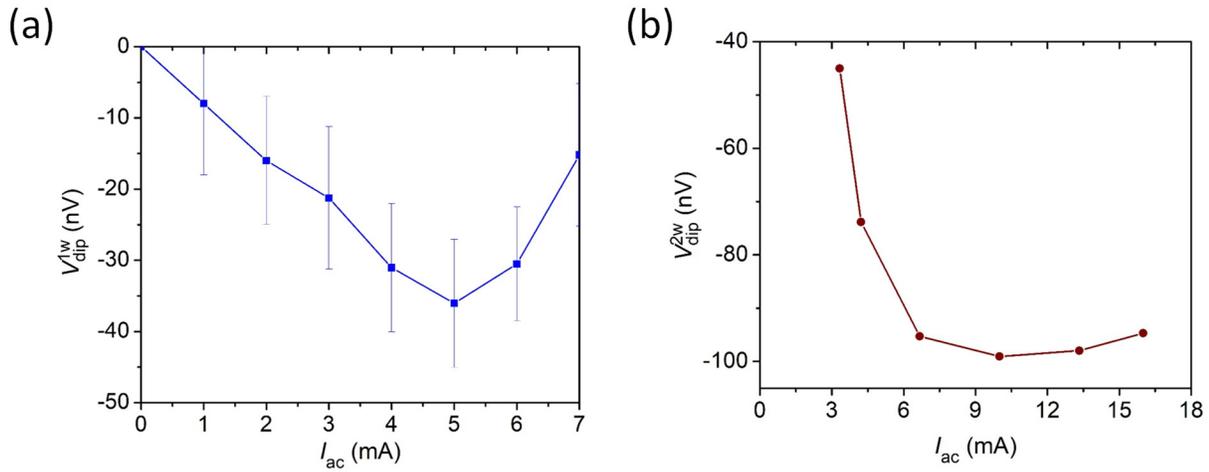

Figure S3. (a) The 1st harmonic nonlocal voltage dip $V_{dip}^{1w}$ measured by the Pt detector at the resonance magnetic field versus the a.c. current applied through the Pt injector, when a constant 4GHz 25dBm microwave is sent into the antenna. (b) The 2nd harmonic nonlocal voltage dip $V_{dip}^{2w}$ measured by the Pt detector at the positive resonance magnetic field versus the a.c. current applied through the Pt injector, when a constant 3GHz 20dBm microwave is sent into the antenna.